\documentclass[11pt]{article}

\usepackage{jheppub}

\usepackage{epsf,epsfig,subfigure,graphicx}

\newcommand{\dis}[1]{\begin{equation}\begin{split}#1\end{split}\end{equation}}
\newcommand{\be}{\begin{equation}}
\newcommand{\ee}{\end{equation}}

\newcommand{\eq}[1]{Eq.~(\ref{#1})}

\def\bea{\begin{eqnarray}}
\def\eea{\end{eqnarray}}

\title{\Large\bf 
Vector-like leptons and extra gauge symmetry 
\\for the natural Higgs boson 
}

\author[a]{Bumseok Kyae,}
\author[b]{Chang Sub Shin}

\affiliation[a]{Department of Physics, Pusan National University, \\Busan 609-735, Korea}
\affiliation[b]{Asia Pacific Center for Theoretical Physics, \\Pohang, Gyeongbuk 790-784, Korea
}

\emailAdd{bkyae@pusan.ac.kr}
\emailAdd{csshin@apctp.org}

\abstract{
For raising the radiative Higgs mass without a serious  fine-tuning in the Higgs sector, we introduce vector-like lepton doublets and neutral singlets 
$\{L,L^c;N,N^c\}$, and consider their order one Yukawa coupling to the Higgs $W\supset y_NLh_uN^c$. 
The 125 GeV Higgs mass can be naturally explained 
with the stop mass squared of $\sim (500~{\rm GeV})^2$ and even without the $A$-term contributions. 
It is possible   
because of the quartic power of $y_N$ in the radiative Higgs mass correction, and much less stringent mass bounds on extra leptonic matter.
In order to avoid blowup of $y_N$ at higher energy scales, a non-Abelian gauge extension of the MSSM is attempted, 
under which $\{L,L^c;N,N^c\}$ are charged, while  
all the ordinary MSSM superfields remain neutral. 
We discuss the gauge coupling unification. 
This mechanism can be applied also for enhancing $h^0\rightarrow\gamma\gamma$ with $W\supset y_EL^ch_uE$, 
if the charged lepton singlets $\{E,E^c\}$ are also introduced.  
} 
\begin{document} 
\maketitle
\flushbottom


\section{Introduction}

The main motivation of introducing supersymmetry (SUSY) at the electroweak (EW) scale is to resolve the gauge hierarchy problem \cite{book}. 
Through SUSY small masses of chiral fermions can be imposed also for bosonic fields like the Higgs in the standard model (SM). 
Thus, the small Higgs mass and the resulting EW scale can be successfully protected against huge quantum corrections. 
As a result, the SM can be successfully embedded in the grand unified theory (GUT) or the string theory. 
The gauge coupling unification in the minimal supersymmetric standard model (MSSM) might be an evidence of such a possibility. 

Recently, ATLAS and CMS collaborations have announced the discovery of the standard model-like Higgs boson around 125 GeV invariant mass \cite{ATLAS,CMS}.   
The Higgs signals at the LHC have seemed to be confirmed in various decay channels of the Higgs boson. 
The Higgs mass of 125 GeV is, however, too heavy to be interpreted as a SUSY Higgs appearing in the MSSM.
It is basically because the tree-level Higgs mass in the MSSM is too small: it is lighter even than the Z boson mass $M_Z$. 
Accordingly, the radiative corrections to it should overcome its smallness, $\Delta m_h^2\gtrsim (86~{\rm GeV})^2$. 

In the MSSM, the radiative Higss mass is induced   
dominantly by the large top quark Yukawa coupling $y_t$ and its corresponding ``$A$-term'' coupling $A_t$ \cite{book,twoloop}: 
\dis{ \label{mssm1}
\Delta m_h^2\approx 
\frac{3v_h^2{\rm sin}^4\beta}{4\pi^2}|y_t|^4 
\left[{\rm log}\left(\frac{m_t^2+\widetilde{m}_t^2}{m_t^2}\right) 
+\left(\frac{X_t}{\widetilde{m}_t}\right)^2
\left\{1-\frac{1}{12}
\left(\frac{X_t}{\widetilde{m}_t}\right)^2\right\}
\right] ,
}
where $v_h$ ($\approx 174$ GeV) indicates the Higgs 
vacuum expectation value (VEV), $m_t^2$ ($\widetilde{m}_t^2$) means the (s)top mass squared, and
 $X_t$ is defined with $A_t$ and the $\mu$ parameter in the MSSM, $X_t\equiv A_t-\mu{\rm cot}\beta$. The last two terms, which come from the $A$-term contribution, are suppressed, unless $(X_t/\widetilde{m}_t)^2\approx 6$ is fulfilled (``maximal mixing scenario''). 
Since $y_t\times {\rm sin}\beta$ and $m_t$ have been already known precisely, only a useful parameter for raising the Higgs mass is the stop mass squared $\widetilde{m}_t^2$. Unfortunately, the logarithmic dependence on $\widetilde{m}_t^2$ in  \eq{mssm1} makes it quite inefficient to raise the radiative Higgs mass. 
For 125 Higgs mass, thus, $\widetilde{m}_t^2$ should be greater than a few TeV at the two-loop level in the MSSM \cite{twoloop}.    

The stop mass squared heavier than $(1 ~{\rm TeV})^2$, however, gives rise to a fine-tuning problem 
as clearly seen from the following one-loop corrected extreme condition for the Higgs scalar potential \cite{book}:   
\dis{ \label{mssm2}
\left\{m_{0}^2+\frac{3|y_t|^2}{8\pi^2}\widetilde{m}_t^2~
{\rm log}\left(
\frac{\widetilde{m}_t^2}{M_G^2}\right)\right\}
+|\mu|^2 
\approx m_3^2{\rm cot}\beta 
+\frac{M_Z^2}{2}{\rm cos}2\beta ,
}
where $m_{0}^2$ denotes the soft mass squared of the u-type Higgs $h_u$ ($\equiv m_2^2$) at the cut-off scale ($=M_G$), and $m_3^2$ stands for the ``$B\mu$'' parameter of the MSSM at the EW scale. 
Here we dropped the $A$-term contributions. 
The first two terms in \eq{mssm2} yield $m_2^2$ at the EW scale. If $\widetilde{m}_t^2$ is excessively large, 
it should be finely tuned with other soft parameters to be equated with the Z boson mass squared of $(91~{\rm GeV})^2$. 

A straightforward resolution to the Higgs mass problem would be to raise the {\it tree} level Higgs mass.  
In the next-to-minimal supersymmetric standard model (NMSSM),
indeed, a singlet $S$ coupled to the Higgs is introduced together with a dimensionless Yukawa coupling $\lambda$ in the superpotential:
\dis{ \label{nmssm}
W_{\rm NMSSM}\supset \lambda S h_uh_d ,
}
which provides an additional quartic Higgs potential, 
remarkably raising the tree level Higgs mass, if $\lambda$ is sizable. 
However, $\lambda$ greater than 0.7 at the EW scale turns out to become blowing-up below the GUT scale through its renormalization group (RG) evolution effect [``Landau-pole (LP) problem''] \cite{Masip}.  
Moreover, for natural explanation of the 125 GeV Higgs mass with avoiding a serious fine-tuning [$(500~{\rm GeV})^2\lesssim\widetilde{m}_t^2\lesssim (700~{\rm GeV})^2$], 
$\lambda$ should be larger than $0.6$ 
with $1\lesssim {\rm tan}\beta \lesssim 3$ \cite{nmssm2}. 
So only a quite narrow band at the edge of the theoretically natural parameter space survives at the moment.   
Hence, relaxing the LP constraint on $\lambda$ is important for the naturalness of the Higgs mass in the NMSSM. 
 
In Ref.~\cite{KS}, an Abelian gauge symmetry, under which $\{h_u,h_d\}$, and $S$ are charged, 
is introduced for relaxing the LP constraint. It is possible because a strong enough new gauge interaction is capable of holding $\lambda$ in the perturbative regime.   
Of course, an asymptotically free non-Abelian gauge symmetry would more effectively work. In this case, however, different flavors of the Higgs should be accompanied. 

An efficient way for raising the {\it radiative} Higgs 
mass is to introduce a new order one Yukawa coupling of unknown vector-like matter and the Higgs \cite{extramatt,extramatt2}. 
Then the new Yukawa coupling would play the role of $y_t$ in \eq{mssm1}. 
Unlike the soft mass parameter, 
the quartic power of a Yukawa coupling
could efficiently enhance the radiative correction, 
if it is sizable. 
However, introduction of new vector-like {\it colored} particle with an order one Yukawa coupling would exceedingly affect the production and decay rates of the Higgs boson at the large hadron collider (LHC). 
Moreover, non-observation of new colored particles so far at the LHC pushes the upper bound on their mass well above 1 TeV. 
Since SUSY mass parameters also appear in \eq{mssm2} together with soft mass squareds, as will be seen later,  
their heavy masses make the fine-tuning problem in the Higgs sector more serious. 
In Ref.~\cite{KP}, thus, 
the possibility that the radiative correction to the Higgs mass is enhanced by the MSSM singlets 
was studied.   
In this paper, we will try to enhance the radiative correction to the Higgs mass with vector-like leptonic matter, since the experimental bounds for extra leptons are not severe yet \cite{PDG,JSW}.

According to recently reported ATLAS data, 
the excess of signal from the SM prediction in the di-photon decay channel of the Higgs is still larger than $2\sigma$ \cite{ATLAS}.
If the deviation will persist even with more analyses and data, 
one of the promising way for explaining it is to introduce 
{\it both} vector-like lepton doublets and singlets, $\{L,L^c;E,E^c\}$ with their sizable Yukawa couplings to the Higgs \cite{C,Bae,JSW},
\dis{ \label{EL}
W_{\rm EL}\supset y_E L^ch_uE + y_E^\prime Lh_dE^c , 
}
Since any new colored particles are not involved in these new Yukawa interactions, 
such a trial leaves intact the production rate of the Higgs at the LHC, and their SUSY mass parameters can be quite small compared to the case of vector-like colored particles. 
As in \eq{nmssm}, however, 
the sizable coupling constants $y_{E,N}$ at the EW scale would diverge below the GUT scale, 
because the strong interaction is not involved also in \eq{EL}.
Actually, since the RG equations governing $y_{E,N}$ (particularly $y_N$) are similar to that of $\lambda$ in the NMSSM, 
the resulting LP constraints are also similar to it, $y_E\lesssim 0.72$ and $y_E^\prime\lesssim 0.94$ at ${\rm tan}\beta =2$ \cite{JSW}.\footnote{The electromagnetic interaction distinguishing 
$E$ and $S$ makes just a small difference of $\sim 0.02$ in the upper bounds of $y_E$ and $\lambda$.  
The absence of the contribution by the top quark Yukawa coupling to the RG equation of $y_E^\prime$ 
admits such a relative relaxation of the LP constraint on $y_E^\prime$ compared to $y_E$.} 
As the data accumulated, moreover, the center value for $h^0\rightarrow\gamma\gamma$ is approaching to the SM prediction with the statistical error decreasing. 
Moreover, the recent CMS report 
says that $\sigma/\sigma_{\rm SM}$ is just $0.8\pm 0.3$ ($1\sigma$) at the moment \cite{CMS}.  

In this paper, we attempt to raise the radiative Higgs mass and 
relieve the fine-tuning in the MSSM by introducing vector-like 
leptons and a new extra gauge symmetry. 
By assuming their relatively light masses and order one Yukawa couplings to the Higgs, 
one can easily enhance the radiative corrections to the Higgs mass, avoiding a serious fine-tuning. 
As mentioned above, however, such order one Yukawa couplings, in which only leptonic particles and the Higgs are involved, would blow up at a high energy scale, unless an extra gauge interaction is supported. 
We assume that only the extra vector-like leptons 
are charged under the extra gauge symmetry. 
Particularly, a non-Abelian extension of the MSSM will be tried, by which the LP can be very easily relaxed.    
It is possible because {\it two} new particles couple to the Higgs unlike the case of \eq{nmssm}.  
This mechanism could be applied also for enhancing $h^0\rightarrow\gamma\gamma$. 
  
This paper is organized as follows:        
in section \ref{sec:radHiggs}, we will discuss the radiative corrections to the Higgs mass and the fine-tuning issue in the presence of the vector-like leptons. 
In section \ref{sec:model}, we will propose a model, introducing an 
extra gauge symmetry, under which the ordinary MSSM superfields including the two Higgs doublets are neutral. 
Section \ref{sec:conclusion} will be devoted to conclusion. 

\section{Radiative Corrections
 } \label{sec:radHiggs}

With the extra vector-like lepton doublets $\{L(n,e),L^c(e^c,n^c)\}$, and the lepton singlets $\{N,N^c, E,E^c\}$, we consider the following superpotential: 
\dis{ \label{superPot}
W =  y_{N}Lh_uN^c + y_EL^ch_uE  
+\mu_L LL^c +   \mu_NNN^c + \mu_E EE^c
+\big\{y_N^\prime L^ch_dN + y_E^\prime Lh_dE^c 
\big\} ,
}
where $y_{N,E}^{(\prime)}$ and $\mu_{L,N,E}$ are  dimensionless and dimensionful parameters, respectively. 
The vector-like leptons acquire their masses from the $\mu_{L,N,E}$ terms. 
The mass bounds for such extra leptons are relatively less severe compared to colored particles at the moment \cite{PDG,JSW}.  
Due to the reason, the fine-tuning is avoidable 
when the extra leptons raise the radiative Higgs mass, 
as will be seen later. 
We will propose an explanation later on why the above $\mu$ parameters are of order the EW scale. 
As mentioned in Introduction, 
the $y_E$, $y_{E}^\prime$ terms in \eq{superPot} can enhance the di-photon decay rate of the Higgs, 
if $y_E$ or $y_{E}^\prime$ is sizable \cite{C}.  
On the other hand, the $y_N$, $y_{N}^\prime$ terms are {\it not} involved in $h^0\rightarrow\gamma\gamma$ at all. 

The $y_{N,E}$ terms in \eq{superPot}, i.e. $h_u$ terms provide radiative corrections to the Higgs mass, 
which are proportional to  ${\rm sin}^4\beta |y_{N,E}|^4$.    
If $y_{N,E}$ are of order unity, hence, they are very helpful for raising the Higgs mass. 
The $h_d$ terms in \eq{superPot} also make contributions to the radiative Higgs mass. 
In contrast to the case of the $h_u$ terms, however, 
they are proportional to ${\rm cos}^4\beta |y_{N,E}^\prime|^4$, 
which is much suppressed for ${\rm tan}\beta\gtrsim 1$. 
Since our prime interest in this paper is to raise the radiative Higgs mass, 
we will neglect the last two terms in \eq{superPot} 
throughout this paper.

From \eq{superPot} one can readily read off the  neutral and charged fermion's mass matrices, ${\cal M}^{\cal F}_{N,E}$. 
In the bases of  $(\overline{L}(\bar{n}),\overline{N};L^c(n^c),N^c)$ and 
$(\overline{L}^c(\bar{e}^c),\overline{E}^c;L(e),E)$, 
the squared mass matrices take the following form:
\begin{eqnarray}
\left({\cal M}^{\cal F}_{N,E}\right)^2 = \left[
\begin{array}{cc|cc}
|\mu_L|^2+|y_{N,E}h_u|^2 & \mu_{N,E}^*y_{N,E}h_u 
& 0 & 0  \\
\mu_{N,E}y_{N,E}^{*}h_u^* & |\mu_{N,E}|^2 & 0 & 0
\\ \hline 
0 & 0 & |\mu_L|^2 & \mu_L^*y_{N,E}h_u  \\
0 & 0 & \mu_Ly_{N,E}^*h_u^* & |\mu_{N,E}|^2+|y_{N,E}h_u|^2
\end{array}\right] . 
\end{eqnarray}
We name the eigenvalues of the $({\cal M}^{\cal F}_{N,E})^2$ for the two heavier degenerate states as  $\left(M^{+}_{N,E}\right)^2$, 
and the eigenvalues for the other two lighter degenerate states as $\left(M^{-}_{N,E}\right)^2$.
$\left(M^{\pm}_{N,E}\right)^2$ are estimated as   
\begin{eqnarray} \label{evalues}
&&\frac12\left[|\mu_L|^2+|\mu_{N,E}|^2+|y_{N,E}h_u|^2
\pm\sqrt{\left(|\mu_L|^2+|\mu_{N,E}|^2+|y_{N,E}h_u|^2\right)^2
-4|\mu_L|^2|\mu_{N,E}|^2}\right] 
\nonumber \\
&&\qquad\quad~~ \approx\left\{
\begin{array}{l}
\vspace{0.2cm}
|\mu_>|_{N,E}^2+
\frac{|\mu_>|_{N,E}^2|y_{N,E}|^2}{|\mu_>|_{N,E}^2-|\mu_<|_{N,E}^2}
|h_u|^2 
\equiv |\mu_>|_{N,E}^2 +|Y^+_{N,E}|^2|h_u|^2
\\
|\mu_<|_{N,E}^2-
\frac{|\mu_<|_{N,E}^2|y_{N,E}|^2}{|\mu_>|_{N,E}^2-|\mu_<|_{N,E}^2}
|h_u|^2
\equiv |\mu_<|_{N,E}^2 -|Y^-_{N,E}|^2|h_u|^2
\end{array}
\right. ,
\end{eqnarray}
where $|\mu_>|_{N,E}^2$ ($|\mu_<|_{N,E}^2$) denotes the heavier 
(lighter) parameter among $|\mu_L|^2$ and $|\mu_{N,E}|^2$.
Just for simplicity, we will assume $|\mu_>|_{N,E}^2\gg |\mu_<|_{N,E}^2$.     
In this limit, we have $|Y^+_{N,E}|^2\approx |y_{N,E}|^2\gg |Y^-_{N,E}|^2$. 
In fact, $|\mu_N|^2$ can be quite smaller than $(100~{\rm GeV})^2$. For the neutral fermions, thus, $|\mu_>|_N^2$ ($|\mu_<|_N^2$) can be $|\mu_L|^2$ ($|\mu_N|^2$).  
In \eq{evalues}, we neglected the quartic corrections $|h_u|^4$, since they are approximately given by 
$\mp 2|y_{N,E}|^4|\mu_<|^2|h_u|^4/|\mu_>|^4$.

The squared mass matrices for the superpartners are 
given by summations of those for the fermions and the SUSY breaking soft mass squareds:  
\dis{
\left({\cal M}^{\cal B}_{N,E}\right)^2=\left({\cal M}^{\cal F}_{N,E}\right)^2 + 
{\rm diag.}(\widetilde{m}_{L,L^c}^2,\widetilde{m}_{N,E^c}^2,\widetilde{m}_{L^c,L}^2,\widetilde{m}_{N^c,E}^2) .
} 
Here we neglected the contributions coming from the $D$- and $A$-terms due to their relative smallness. 
Note that the contributions by the MSSM $\mu$ term coming from the cross terms of $|\partial W/\partial h_u|^2$ are ${\rm tan}\beta$-suppressed, since they are proportional to $h_d$.
For simplicity, we set all the soft mass squareds equal to $\widetilde{m}^2$ in this paper. 
The mass differences between the bosonic and fermionic modes of the vector-like leptonic superfields could induce non-zero radiative corrections to the Higgs potential.  

After integrating out the heavy fields associated with  $\left({\cal M}^{\cal F}_{N,E}\right)^2$ and $\left({\cal M}^{\cal B}_{N,E}\right)^2$, 
one can get the one-loop effective scalar potential for the light field \cite{CW,book}, i.e. the Higgs boson in this case: 
\dis{ \label{CWpot}
&\qquad\quad \Delta V=\frac{1}{32\pi^2}{\rm Tr}\bigg[{\cal M_B}^4\left\{{\rm log}\frac{{\cal M_B}^2}{Q^2}-\frac32\right\}
-{\cal M_F}^4\left\{{\rm log}\frac{{\cal M_F}^2}{Q^2}-\frac32\right\}\bigg]
\\
&\approx\frac{1}{16\pi^2}\sum_{i}\bigg[( M_i^2+\widetilde{m}^2)^2\left\{{\rm log}\left(\frac{M_i^2+\widetilde{m}^2}{Q^2}\right)-\frac32\right\}
-(M_i^2)^2\left\{{\rm log}\left(\frac{M_i^2}{Q^2}\right)-\frac32\right\}\bigg] ,
}
where ${\cal M_B}^2$ (${\cal M_F}^2$) indicates 
the squared mass matrix for bosonic (fermionic) modes,
and $M_i^2=\{(M^+_N)^2, (M^-_N)^2, (M^+_E)^2, (M^-_E)^2\}$. $Q$ denotes the renormalization scale.  
Here we ignored the $A$-term contributions. 
The radiative correction to the Higgs potential \eq{CWpot} is expanded in powers of $|\delta h_u|^2$ ($\equiv |h_u|^2-v_h^2{\rm sin}^2\beta$) as follows:  
\dis{ \label{expansion}
\Delta V\approx \Delta V_0+\left(\partial_{h_u}\partial_{h_u^*}\Delta V\right)_0 |\delta h_u|^2+\frac{1}{2!2!}\left(\partial^2_{h_u}\partial^2_{h_u^*}\Delta V\right)_0 |\delta h_u|^4+\cdots  
}
The coefficients of the quadratic 
and quartic terms of $\delta h_u$ in \eq{expansion}  are estimated as 
\begin{eqnarray} \label{coeff}
\left(\partial_{h_u}\partial_{h_u^*}\Delta V\right)_0&\approx& \sum_{i=N,E}
\frac{|y_i|^2}{8\pi^2}
\left[\left(\overline{M}_i^2+\widetilde{m}^2\right)\left\{{\rm log}\left(\frac{\overline{M}_i^2+\widetilde{m}^2}{Q^2}\right)-1\right\}
-\overline{M}_i^2\left\{{\rm log}\left(\frac{\overline{M}_i^2}{Q^2}\right)-1\right\}\right]
\nonumber \\
&\equiv& \sum_{i=N,E}\frac{|y_i|^2}{8\pi^2}
\bigg[f_Q(\overline{M}_i^2+\widetilde{m}^2)-f_Q(\overline{M}_i^2)\bigg] ,
\\
\left(\partial^2_{h_u}\partial^2_{h_u^*}\Delta V\right)_0&\approx&\sum_{i=N,E}\frac{|y_i|^4}{4\pi^2}~ {\rm log}\left(\frac{\overline{M}_i^2
+\widetilde{m}^2}{\overline{M}_i^2}\right) ,
\nonumber
\end{eqnarray}
where $\overline{M}_i^2$ and $f_Q(m^2)$ are defined as $\overline{M}_i^2\equiv |\mu_>|^2+|y_i|^2v_h^2{\rm sin}^2\beta$ and $f_Q(m^2)\equiv m^2\{{\rm log}(\frac{m^2}{Q^2})-1\}$, respectively. 
Note that if we have $N_V$ copies of $\{L,L^c;N,N^c;E,E^c\}$ 
{\it by a symmetry}, \eq{coeff} should be multiplied by $N_V$, and so 
$|y_i|^2$ and $|y_i|^4$ in \eq{coeff} are replaced by $N_V|y_i|^2$ and $N_V|y_i|^4$, respectively. We will consider this possibility later.  

The quadratic term in \eq{expansion}, 
which depends on $Q$, renormalizes the soft mass squared of the u-type Higgs in the MSSM, $m_{2}^2(Q)$
together with the (s)top contribution:  
\dis{ \label{renorm}
m_2^2(Q)+\frac{3|y_t|^2}{8\pi^2}
\bigg[f_Q(m_t^2+\widetilde{m}_t^2)-f_Q(m_t^2)\bigg]
+\sum_{i=N,E}\frac{|y_i|^2}{8\pi^2}
\bigg[f_Q(\overline{M}_i^2+\widetilde{m}^2)-f_Q(\overline{M}_i^2)\bigg] .
}
Inserting the RG solution of $m_2^2(Q)$ into \eq{renorm} replaces the $Q$ dependence in \eq{renorm} by the cut-off scale, 
in which the soft parameters are generated, 
yielding the low energy value of $m_2^2$ \cite{CQW}. 
In the minimal SUGRA, the GUT scale is adopted as the cut-off scale, and so  
\dis{
m_2^2|_{\rm EW}\approx m_0^2+ \frac{3|y_t|^2}{8\pi^2}\widetilde{m}_t^2~
{\rm log}\left(
\frac{\widetilde{m}_t^2}{M_G^2}\right) 
+\sum_{i=N,E}\frac{|y_i|^2}{8\pi^2}
\bigg[f_Q(\overline{M}_i^2+\widetilde{m}^2)-f_Q(\overline{M}_i^2)\bigg]_{Q=M_G} , 
}
where $m_0^2$ stands for the value of $m_2^2$ at the GUT scale.

One of the extremum conditions in the Higgs potential ($\partial_{h_u}V_H=\partial_{h_d}V_H=0$) is \cite{book,twoloop}
\dis{ \label{extremum}
m_2^2|_{\rm EW}+|\mu|_{\rm EW}^2 
\approx m_3^2|_{\rm EW}~{\rm cot}\beta 
+\frac{M_Z^2}{2}{\rm cos}2\beta ,
}
which should, of course, be fulfilled around the vacuum state. 
Indeed, $M_Z^2$ [$=(g_2^2+g_Y^2)v_h^2/2\approx(91~ {\rm GeV})^2$] defines the EW scale. 
For the naturalness of the EW scale and its perturbative stability, $\overline{M}_i^2$,  $\widetilde{m}^2$ and $\widetilde{m}_t^2$ 
should be much smaller than (1 TeV)$^2$.
Otherwise, the mass parameters appearing in \eq{extremum} should be finely tuned. 
Unlike vector-like extra {\it colored} particles, the mass parameters associated with the vector-like extra {\it leptons} are not severely constrained from the LHC data: they could be much lighter than 500 GeV \cite{PDG,JSW}.      
In this paper we suppose that 
\dis{ \label{naturalpara}
\widetilde{m}_t^2 \gtrsim (500~{\rm GeV})^2 ~; ~~
(100 ~{\rm GeV})^2~~ \lesssim ~~ 
|\mu_L|^2 , ~ |\mu_E|^2 ,~  \widetilde{m}^2 ~~ 
\lesssim ~~ (600 ~{\rm GeV})^2 ~, 
}
while $|\mu_N|^2$ can be even smaller than $(100~{\rm GeV})^2$. They can avoid the LEP bound \cite{PDG}.

The quartic term in \eq{expansion}, 
which is independent of the renormalization scale $Q$, 
contributes to the radiative correction to the Higgs mass together with the (s)top \cite{book}: 
\dis{
m_h^2\approx M_Z^2{\rm cos^22\beta}+\Delta m_h^2|_{\rm top} + \Delta m_h^2|_{\rm NE} ~\approx ~ (125~{\rm GeV})^2 ,
}
where the (s)top and vector-like leptons' contributions, $\Delta m_h^2|_{\rm top}$ and $\Delta m_h^2|_{\rm NE}$
are presented as follows: 
\dis{
&\Delta m_h^2|_{\rm top}\approx \frac{3v_h^2{\rm sin}^4\beta}{4\pi^2}|y_t|^4 ~{\rm log}\left(\frac{m_t^2+\widetilde{m}_t^2}{m_t^2}\right)
= \frac{3m_t^4}{4\pi^2v_h^2} ~{\rm log}\left(\frac{m_t^2+\widetilde{m}_t^2}{m_t^2}\right),
\\
&~~ \Delta m_h^2|_{\rm NE}\approx \frac{v_h^2{\rm sin}^4\beta}{4\pi^2}
\sum_{i=N,E}|y_i|^4
~{\rm log}\left[\frac{|\mu_>|_i^2+|y_i|^2v_h^2{\rm sin}^2\beta+\widetilde{m}^2}{|\mu_>|_i^2+|y_i|^2v_h^2{\rm sin}^2\beta}\right] .
}
For $m_t^2\approx (173~{\rm GeV})^2$ and $\widetilde{m}_t^2\gtrsim (500~{\rm GeV})^2$, 
we have just $\Delta m_h^2|_{\rm top}\gtrsim (70.9~{\rm GeV})^2$. 
For explaining the observed 125 GeV Higgs mass, 
thus, it is required that   
%
\dis{ \label{requirement}
N_V\sum_{i=N,E}|y_i|^4 
~{\rm log}\left[\frac{|\mu_>|_i^2+|y_i|^2v_h^2{\rm sin}^2\beta+\widetilde{m}^2}{|\mu_>|_i^2+|y_i|^2v_h^2{\rm sin}^2\beta}\right] ~ \lesssim ~  15.5,~6.1,~4.4,~3.5,~3.0    
}
for ${\rm tan}\beta=2,4,6,10,50$, respectively. 
Here we add the factor ``$N_V$'' 
in order to include also cases with $N_V$ copies of $\{L,L^c;N,N^c;E,E^c\}$ by a symmetry.
Of course, $N_V=1$ for \eq{superPot}. 
Note that for $y_{E}\approx 0.7$ (so $|y_E|^4\approx 0.24$), which is the maximal value allowed at the EW scale, avoiding the LP constraints \cite{JSW}, the logarithmic part in \eq{requirement} should be in the range of 12.5 (for ${\rm tan}\beta=50$) -- 64.6 (for ${\rm tan}\beta=2$). 
On the contrary, the logarithmic part in the parameter space of \eq{naturalpara} is smaller than 3.6 for $\widetilde{m}^2=(600~{\rm GeV})^2$.  
Actually, a similar LP constraint is applicable to $y_N$, because of the similarity in the RG equations of $y_E$ and $y_N$.  
It means that the observed Higgs mass is impossible to be explained in the parameter space of \eq{naturalpara}. We note, however, that even slight relaxation of such LP constraints 
would be very helpful, if it is somehow possible, because of the relative high power of $y_{N,E}$ in \eq{requirement}. For instance, if $|y_N|=1.5$ (so $|y_N|^4\approx 5$) is somehow permitted, 
\eq{requirement} is easily satisfied with \eq{naturalpara} for ${\rm tan}\beta\gtrsim 3$.  
We will see that $y_N$ can reach 1.78 or 2.0 (so $|y_N|^4\approx 10$ or $16$) by introducing an extra SU(2) gauge symmetry and more matter.   


\section{The Model}
\label{sec:model}

In this section, we attempt to relax the LP constraints 
on $y_{N,E}$ by introducing an extra SU(2)$_{Z'}$ {\it or} U(1)$_{Z'}$ {\it gauge} symmetry. 
Under the SU(2)$_{Z'}$ [U(1)$_{Z'}$], 
the extra vector-like leptons $\{L,L^c,E,E^c,N,N^{c},N_H,N_H^c\}$ are assumed to be the fundamental representations [charged], 
whereas {\it the ordinary matter in the MSSM including the two Higgs doublets are all neutral}.  
The local and global quantum numbers of them are listed in Table 1. All the gauge anomalies with the field contents in Table 1 are free. 
\begin{table}[!h]
\begin{center}
\begin{tabular}
{c|cccc|cccc|c} 
{\rm Superfields}  &   ~$L$~   &
 ~$L^c$~  &  ~$E$~ & ~$E^c$~ & ~$N$~  &  ~$N^c$~ & ~$N_H$ & ~$N_H^{c}$~ & ~$X$
  \\
\hline 
SU(2)$_{Z'}$ [U(1)$_{Z'}$]  & ~${\bf 2}$$[1]$ & ${\bf 2}$$[-1]$ & ${\bf 2}$$[1]$ & ${\bf 2}$$[-1]$ & ~${\bf 2}$$[1]$ & ${\bf 2}$$[-1]$
 & ${\bf 2}$$[1]$ & ${\bf 2}$$[-1]$ & ~~${\bf 1}$$[0]$  
 \\
 U(1)$_{\rm R}$  & ~$1$ & ~$1$ & ~$1$ & ~$1$ & ~$1$ & ~$1$ 
 & ~$0$ & ~$2$  & ~~$2$ 
 \\
 U(1)$_{\rm PQ}$  & $-1$ & $-1$ & ~$1$ & $-3$ & $-3$ & ~$1$ & $-1$ & $-1$
 & $-2$
\end{tabular}
\end{center}\caption{Matter fields charged under the gauge SU(2)$_{Z'}$ [U(1)$_{Z'}$] and/or  
the global U(1)$_{\rm R}\times$U(1)$_{\rm PQ}$ symmetries.
The ordinary superfields of the MSSM are all {\it inert} under SU(2)$_{Z'}$ [U(1)$_{Z'}$]. Here we dropped the U(1)$_{Z'}$ charge normalization $1/\sqrt{N_Q}$.
}\label{tab:Qnumb1}
\end{table}

The relevant superpotential and 
the K${\rm \ddot{a}}$hler potential are 
\dis{ \label{WK}
&\qquad\qquad\qquad~~ W = y_N Lh_uN^c + y_E L^ch_uE ,
\\
& K=\frac{X^\dagger}{M_P}\left(\kappa_LLL^c
+\kappa_EEE^c+\kappa_NNN^c+\kappa_HN_HN_H^c
\right) + {\rm h.c.} ,
}
where $X$ denotes a spurion superfield for SUSY breaking effects: its F-component develops a VEV of order $\sqrt{m_{3/2}M_P}$, breaking U(1)$_{\rm PQ}$.
U(1)$_{\rm R}$ is broken to the $Z_2$ symmetry by the instanton effects, the $A$-terms in the scalar potential, etc.,  
which can be identified with the matter parity in the MSSM. 
From the K${\rm \ddot{a}}$hler potential, 
the $\mu_{L,N,E}$ terms in \eq{superPot}, 
and also the ``$\mu$ term'' for $\{N_H,N_H^c\}$ 
can be generated with the desired sizes \cite{GM}.
Their magnitudes can be controlled with the $\kappa$s.
$\{N_H,N_H^c\}$ play the role of the ``Higgs'' breaking SU(2)$_{Z'}$ [or U(1)$_{Z'}$]. 
%
$\{N_H,N_H^c\}$ can couple to the hidden sector fields with order one Yukawa couplings, which are not specified in this paper. 
Then, the soft mass squareds of $\{N_H,N_H^c\}$ could become negative at low energies via the RG running 
as in the MSSM Higgs, breaking SU(2)$_{Z'}$ [or U(1)$_{Z'}$] completely. 
Since the SU(2)$_{Z'}$ sector fields are basically leptonic or neutral under the MSSM, 
the spontaneous breaking scale can be much below 1 TeV. However, we take a conservative value: we suppose it is around 1 TeV. 
Since the SU(2)$_{Z'}$ breaking sector is completely separated from the MSSM Higgs, it does not affect the radiative Higgs mass correction. 

In this paper, we intend to maintain the MSSM gauge coupling unification.
Thus, we will study the following cases: 

\vspace{0.3cm}

\underline{\bf Case I}: One pair of $\{L,L^c\}$, 
which are SU(2)$_{Z'}$ doublets and 
two pairs of heavy SU(3)$_c$ triplets, $\{D,D^c\}$, 
which are SU(2)$_{Z'}$ singlets, are essentially present. 
They compose $2\times \{{\bf 5},\overline{\bf 5}\}$ of SU(5). 
The SU(2)$_{Z'}$ doublets, $\{N,N^c;N_H,N_H^c\}$ should also essentially exist. 
In Case I, hence, $N_V$ in \eq{requirement} is given by $2$.
If necessary, one can introduce more $\{{\bf 5},\overline{\bf 5}\}$ of SU(2)$_{Z'}$ singlets, or  $\{N,N^c\}$ pairs of SU(2)$_{Z'}$ doublets with their mass terms in the superpotential. 
Of course, U(1)$_{Z'}$ is also available. However, we will not consider it in this case for simplicity. 
Since $\{E,E^c\}$ are absent, the $y_E$ term in \eq{WK} is ignored in Case I.   

\vspace{0.3cm}

\underline{\bf Case II}: Introduction of one pair of $\{L,L^c;E,E^c\}$ needs to be supplemented 
by $\{Q,Q^c;U,U^c;D,D^c\}$ in order to compose the SU(5) multiplets, 
$\{{\bf 10},\overline{\bf 10}\}$ and $\{{\bf 5},\overline{\bf 5}\}$. One pair of $\{{\bf 10},\overline{\bf 10};{\bf 5},\overline{\bf 5}\}$ makes already the SU(3)$_c$ gauge coupling, $g_3$ increasing at higher energies. 
Accordingly, introduction of SU(2)$_{Z'}$ with $\{L,L^c;E,E^c\}$, which eventually requires at least $2\times\{{\bf 10},\overline{\bf 10};{\bf 5},\overline{\bf 5}\}$, 
results in blowup of the MSSM gauge couplings below the GUT scale. In Case II, hence, we consider only U(1)$_{Z'}$, and the matter contents of only one pair of $\{L,L^c;E,E^c;N_H,N_H^c\}$ plus one pair of heavy $\{Q,Q^c;U,U^c;D,D^c\}$. Hence, $N_V=1$ in Case II.  
While $\{L,L^c;E,E^c;N_H,N_H^c\}$ carry the U(1)$_{Z'}$ charges of $\pm 1$, as displayed in Table 1,   $\{Q,Q^c;U,U^c;D,D^c\}$ are neutral under U(1)$_{Z'}$.  
For simplicity, we ignore $\{N,N^c\}$ in this case. 
The absence of $\{N,N^c\}$ allows us to neglect the $y_N$ term in \eq{WK}.        

\vspace{0.3cm}

Considering the gauge quantum numbers in Table 1 and the Yukawa interactions in \eq{WK}, we list below   
the one-loop anomalous dimensions for the extra vector-like leptons, the MSSM u-type Higgs, 
and the third generation of the quarks, $q_3$ and  $u_3^c$: 
\begin{eqnarray} \label{gammaI}
&&{\rm Case~ I}~~~~\left\{
\begin{array}{l}
\vspace{0.2cm}
16\pi^2\gamma^{L}_{L}=|y_N|^2
-\frac32g_2^2-\frac{3}{10}g_1^2-\frac32g_{Z'}^2 ,
\\ \vspace{0.2cm}
16\pi^2\gamma^{N^c}_{N^c}=2|y_N|^2
-\frac32g_{Z'}^2 ,
\\ 
16\pi^2\gamma^{h_u}_{h_u}=2|y_N|^2 + 
3|y_t|^2-\frac32 g_2^2-\frac{3}{10}g_1^2 ,
\end{array}
\right. 
%
\\ \label{gammaII} 
&&{\rm Case~ II}~~~\left\{
\begin{array}{l}
\vspace{0.2cm}
16\pi^2\gamma^{L^c}_{L^c}=|y_E|^2
-\frac32g_2^2-\frac{3}{10}g_1^2-\frac{2}{N_Q}g_{Z'}^2 ,
\\ \vspace{0.2cm}
16\pi^2\gamma^{E}_{E}=2|y_E|^2-\frac65g_1^2
-\frac{2}{N_Q}g_{Z'}^2 ,
\\ 
16\pi^2\gamma^{h_u}_{h_u}= |y_E|^2 + 
3|y_t|^2-\frac32 g_2^2-\frac{3}{10}g_1^2 ,
\end{array}
\right. 
\\ \label{gamma}
%
&&{\rm MSSM}~~~~\left\{
\begin{array}{l}
\vspace{0.2cm} 
16\pi^2\gamma^{q_3}_{q_3}=|y_t|^2
-\frac83g_3^2-\frac32g_2^2-\frac{1}{30}g_1^2 ,
\\
16\pi^2\gamma^{u_3^c}_{u_3^c}=2|y_t|^2
-\frac83g_3^2-\frac{8}{15}g_1^2 , 
\end{array}
\right. 
\end{eqnarray}
where $y_t$ means the top quark Yukawa coupling. 
In $\gamma^{q_3}_{q_3}$ of \eq{gamma}, we neglected the bottom quark's Yukawa coupling due to the relative smallness for ${\rm tan}\beta\lesssim 16.7$.
The $g_{3,2,1}^2$ terms in Eqs.~(\ref{gammaI}), (\ref{gammaII}), and (\ref{gamma}) originate from the MSSM gauge interactions.  
Since we have an additional SU(2)$_{Z'}$ [or U(1)$_{Z'}$] gauge interaction, the terms of $g_{Z'}^2$ also appear in Eqs.~(\ref{gammaI}) and  (\ref{gammaII}). 
Such MSSM and extra gauge interactions make the {\it negative} contributions to the anomalous dimensions.  

With Eqs.~(\ref{gammaI}), (\ref{gammaII}), and (\ref{gamma}) one can readily write down the RG equations for $y_N$, $y_E$, and also the top quark Yukawa coupling $y_t$: 
\begin{eqnarray} \label{RGeqI}
&&{\rm Case~ I}~~~~\left\{
\begin{array}{l}
\vspace{0.3cm}
\frac{d|y_N|^2}{dt}= \frac{|y_N|^2}{8\pi^2}\bigg[5|y_N|^2+3|y_t|^2-3g_2^2-\frac35g_1^2-3g_{Z'}^2
\bigg] ,
\\
\frac{d|y_t|^2}{dt}=\frac{|y_t|^2}{8\pi^2}\bigg[2|y_N|^2+6|y_t|^2-\frac{16}{3}g_3^2-3g_2^2-\frac{13}{15}g_1^2
\bigg] ,
\end{array}
\right. 
\\ \label{RGeqII}
&&{\rm Case~ II}~~~\left\{
\begin{array}{l}
\vspace{0.3cm}
\frac{d|y_E|^2}{dt}= \frac{|y_E|^2}{8\pi^2}\bigg[4|y_E|^2+3|y_t|^2-3g_2^2-\frac95g_1^2-\frac{4}{N_Q}g_{Z'}^2
\bigg] ,
\\
\frac{d|y_t|^2}{dt}=\frac{|y_t|^2}{8\pi^2}\bigg[|y_E|^2+6|y_t|^2-\frac{16}{3}g_3^2-3g_2^2-\frac{13}{15}g_1^2
\bigg] .
\end{array}
\right. 
%
\end{eqnarray}
From Eqs.~(\ref{RGeqI}) and (\ref{RGeqII}), we can expect that the LP constraint can be remarkably relaxed by the additional negative contributions coming from $g_{Z'}^2$ terms. 
As a result, the allowed maximal values for $y_{N,E}$ 
can be lifted up, compared to the case that the extra gauge symmetry is absent.
   
We note that the RG equations of \eq{RGeqII} were the same as those associated with the Yukawa coupling ``$\lambda Sh_uh_d$'' in the NMSSM except for the Abelian gauge interactions. Actually $g_1^2$ is quite small, and still remains small even up to the GUT scale. 
In the absence of U(1)$_{Z'}$, thus, 
the maximally allowed $y_E$ at the EW scale would be a similar value to that of $\lambda$ in the NMSSM, i.e. 0.7 \cite{Masip}.  
Moreover, were it not for the SU(2)$_{Z'}$ gauge interaction, the RG equations of \eq{RGeqI} became  exactly coincident with those associated with $\lambda$ of the NMSSM. 
Hence, we can get the same upper bound for $y_N$, 
i.e. $|y_N|\lesssim 0.7$ for avoiding the LP constraint in the absence of SU(2)$_{Z'}$.   


The three MSSM gauge couplings in Eqs.~(\ref{RGeqI}) and (\ref{RGeqII}) are given by 
\begin{eqnarray} \label{RGgauge}
g_k^2(t)=\frac{g_U^2}
{1+\frac{g_U^2}{8\pi^2}b_k(t_0-t)} \qquad {\rm for} ~~k=3,~2,~1 ,
\end{eqnarray}
where $t$ parametrizes the renormalization scale, 
$t-t_0={\rm log}(Q/M_{\rm GUT})$. $b_k$ ($k=3,2,1$) denotes the beta function coefficients of the gauge couplings for SU(3)$_c$, SU(2)$_L$ and U(1)$_Y$ [with the SU(5) normalization]. 
In the existence of the extra $v$ pairs of $\{{\bf 5},\overline{\bf 5}\}$, 
they are given by $b_k=(-3+v,1+v,33/5+v)$. 
$v=0$ corresponds to the case of the MSSM.
For the matter contents of Table \ref{tab:Qnumb1} ($v=2$) in Case I, the unified gauge coupling $g_U^2$ is estimated as $0.82$.
%
For $v=3$, $4$, and $5$, $g_U^2$ is lifted to $1.18$, $2.13$, and $11.19$, respectively.
On the other hand, Case II corresponds effectively to the case of $v=4$, and so $g_U^2$ is given by $2.13$. 
In fact, relatively heavier colored particles can cure the small deviation of the gauge coupling unification at the two loop level. In this paper, however, we ignore it.  

Similar to \eq{RGgauge}, the solution to the RG equation of the extra SU(2)$_{Z'}$ [U(1)$_{Z'}$] gauge coupling is 
\begin{eqnarray}  \label{SU2gauge}
g_{Z'}^2(t)=\frac{g_{Z'0}^2}
{1+\frac{g_{Z'0}^2}{8\pi^2}b_{Z'}(t_0-t)} ~~ {\rm for}~~ t>t_{Z'}   
\end{eqnarray}
where $b_{Z'}$ indicates the beta function coefficient of SU(2)$_{Z'}$ or U(1)$_{Z'}$, 
and $t_{Z'}$ parametrizes the SU(2)$_{Z'}$ or U(1)$_{Z'}$ breaking scale $M_{Z'}$ 
[$t_{Z'}-t_0\equiv {\rm log}(M_{Z'}/M_{\rm GUT})$].
In Case I, we have $\{L,L^c;N,N^c;N_H,N_H^c\}$ and extra $n_{N}\times\{N,N^c\}$, and the extra gauge group is SU(2)$_{Z'}$. So $b_{Z'}=-2+n_{N}$. 
In Case II, there are $\{L,L^c;E,E^c;N_H,N_H^c\}$, but the extra gauge group is just U(1)$_{Z'}$, which gives $b_{Z'}=8/N_Q$, where $N_Q$ denotes the charge normalization. 
Unless U(1)$_{Z'}$ is embedded in a simple group, 
the normalization factor $N_Q$ remains undetermined.

\begin{figure}
\begin{center}
\subfigure[]
{\includegraphics[width=0.46\linewidth]{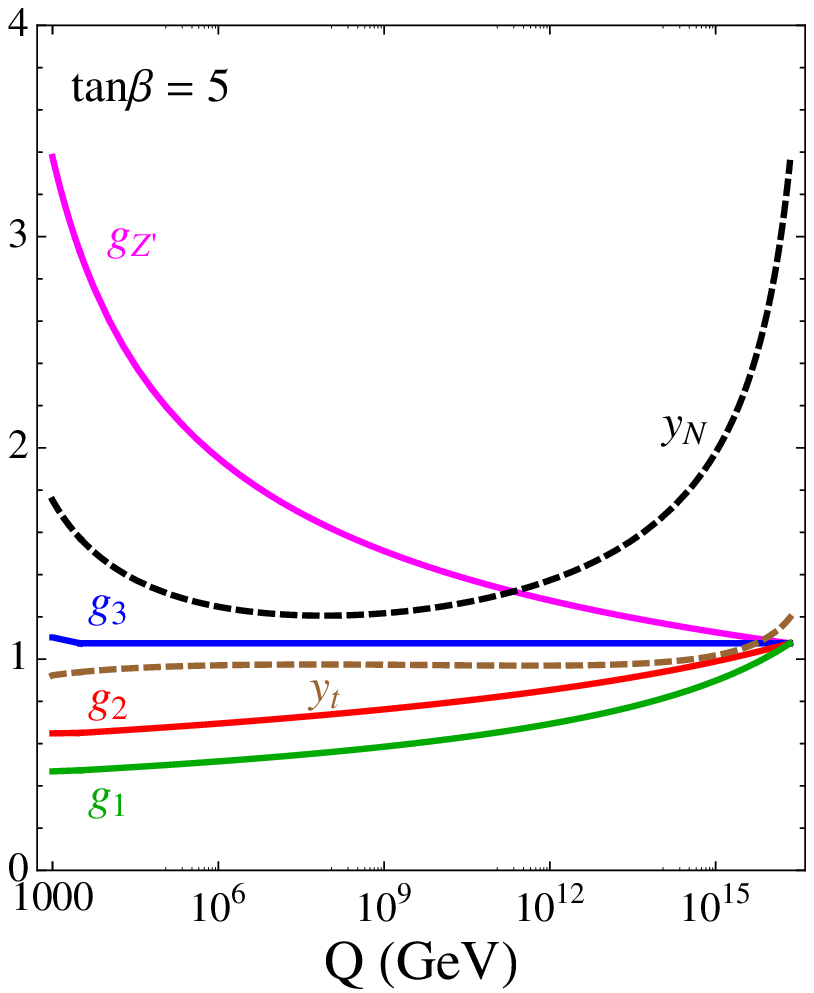}}
\hspace{0.2cm}
\subfigure[] 
{\includegraphics[width=0.51\linewidth]{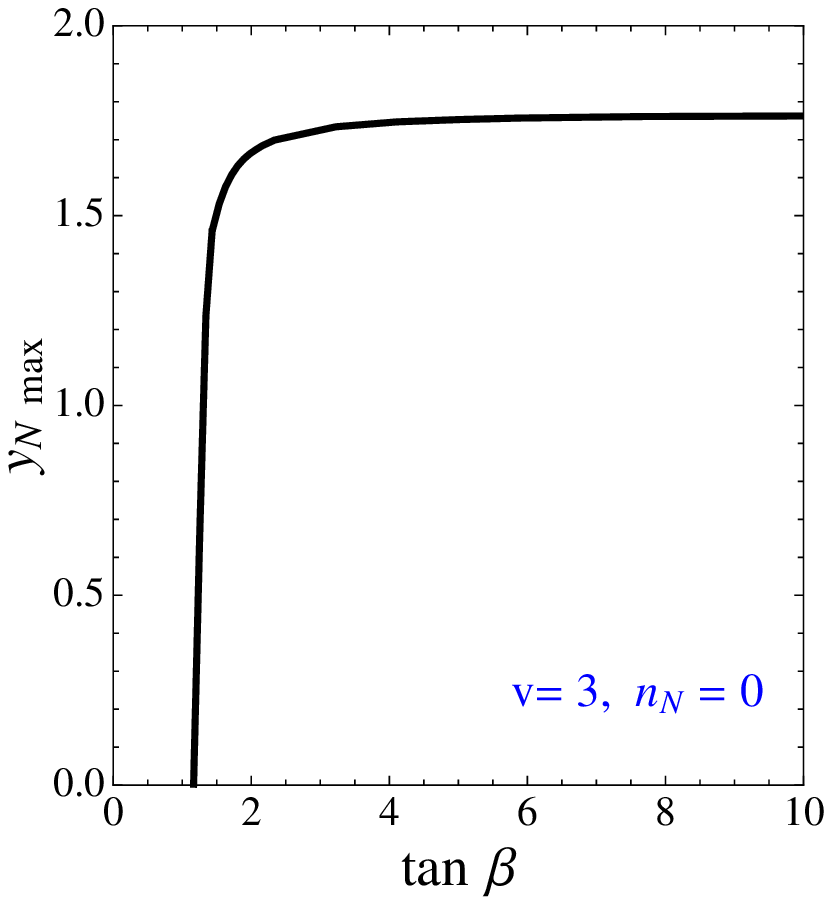}}
\end{center}
\caption{(a) RG runnings of the various couplings 
for the SU(2)$_{Z'}$ extension (Case I).
The MSSM gauge couplings $\{g_3,g_2,g_1\}$ and the SU(2)$_{Z'}$ gauge coupling $g_{Z'}$ are unified 
at the GUT scale ($=2\times 10^{16}~{\rm GeV}$). 
SU(2)$_{Z'}$ is spontaneously broken around 1 TeV. 
(b) $y_{N \rm max}$ vs. tan$\beta$ for the SU(2)$_{Z'}$ extension (Case I). 
With $v=3$ and $n_N=0$ the upper bound on $y_N$ at the EW scale, $y_{N \rm max}$ is lifted 
to 1.78 for ${\rm tan}\beta\gtrsim 3$. 
In a similar way, $y_{N\rm max}$ can reach 2.0 with $v=4$ and $n_N=1$.   
}
\end{figure}

In Case I, as seen from Figure 1, the maximal value of $y_N$ at low energy is lifted to 1.78 (so $|y_N|^4\approx 10$) for $v=3$,  $n_N=0$, and ${\rm tan}\beta\gtrsim 3$ with all the gauge couplings including $g_{Z'}$ unified at the GUT scale ($=2\times 10^{16}~{\rm GeV}$). The top quark Yukawa coupling also become similar to the unified gauge couplings at the GUT scale. 
Here the masses of $\{D,D^c\}$ are set to be 3 TeV. 
Hence, this model can potentially be embedded in a (higher dimensional) unified theory like string theory. However, we don't specify what it is in this paper. 
Since \eq{requirement} is very easily satisfied for ${\rm tan}\beta\gtrsim 2$, the 125 GeV Higgs mass is explained without a serious fine-tuning.   
Particularly, if we set $\widetilde{m}^2=|\mu_L|^2$ ($\gg v_h^2$), then \eq{requirement} becomes simplified to $2\times |y_N|^4\approx 22.4$, $8.8$, $6.3$, $5.0$, $4.3$ for ${\rm tan}\beta=2$, 4, 6, 10, 50. Note that $N_V=2$ in Case I. 
In this case, thus, 125 GeV Higgs mass can be explained as long as ${\rm tan}\beta\gtrsim 2-3$.
In a similar way, we have achieved $y_N=2.0$ at low energy for $v=4$ and $n_N=1$.

Figure 1 shows that $g_{Z^\prime}$ reaches $3.4$ 
(and so the expansion parameter, $g_{Z^\prime}^2/4\pi$ becomes $0.9$) at 1 TeV, which is almost the maximal value of $g_{Z^\prime}$ that the perturbativity allows. 
In this case, $y_N$ reaches the perturbativity bound ($\approx 3.4$) at the GUT scale. 
We consider such an extreme case in order to obtain the upper bound of $y_N$ at low energy, which is shown in Figure 1-(b).  
As discussed above, $y_N$ needed for explaining the 125 GeV Higgs mass can be quite smaller than the upper bound, $y_{N{\rm max}}=1.78$ for ${\rm tan}\gtrsim 3$. 
Although a smaller value of $g_{Z^\prime}$ was taken, 
the gauge coupling unification can still be achieved 
with threshold corrections by heavy matter around the GUT scale. 
However, one should note that SU(2)$_{Z^\prime}$ is  spontaneously broken around 1 TeV energy scale 
by the non-zero VEVs, $\langle N_H\rangle$ and $\langle N_H^c\rangle$, as explained already.  

The sizable Yukawa coupling $y_N$ can affect the oblique parameters $T$ and $S$. 
The experimental best fit for $(\Delta T,\Delta S)$ with respect to the SM reference requires that  
$0.01\lesssim\Delta S\lesssim 0.17$ (1$\sigma$) for $\Delta T\approx 0.12$, and $m_h = 125.7\pm 0.4 ~{\rm GeV}$, $m_t = 173.18\pm 0.94 ~{\rm GeV}$ \cite{STU}.  
$\Delta T\approx 0.12$ constrains the parameter space as \cite{extramatt2} 
\dis{ \label{obliqueT}
N_V\times |y_N|^4\left(
\frac{500~{\rm GeV}}{|\mu_L|}\right)^2
{\rm sin}^4\beta \approx 5.56 .
}
For $\widetilde{m}^2=|\mu_L|^2$ ($\gg v_h^2$),
thus, \eq{obliqueT} provides the lower mass bounds on $|\mu_L|$; 803 GeV, 592 GeV, 517 GeV, 469 GeV, and 440 GeV in Case I for ${\rm tan}\beta=2$, 4, 6, 10, and 50, respectively.
In this parameter range, $\Delta S$ turns out to be $0.01\lesssim\Delta S\lesssim 0.02$.

\begin{figure}
\begin{center}
\subfigure[]
{\includegraphics[width=0.46\linewidth]{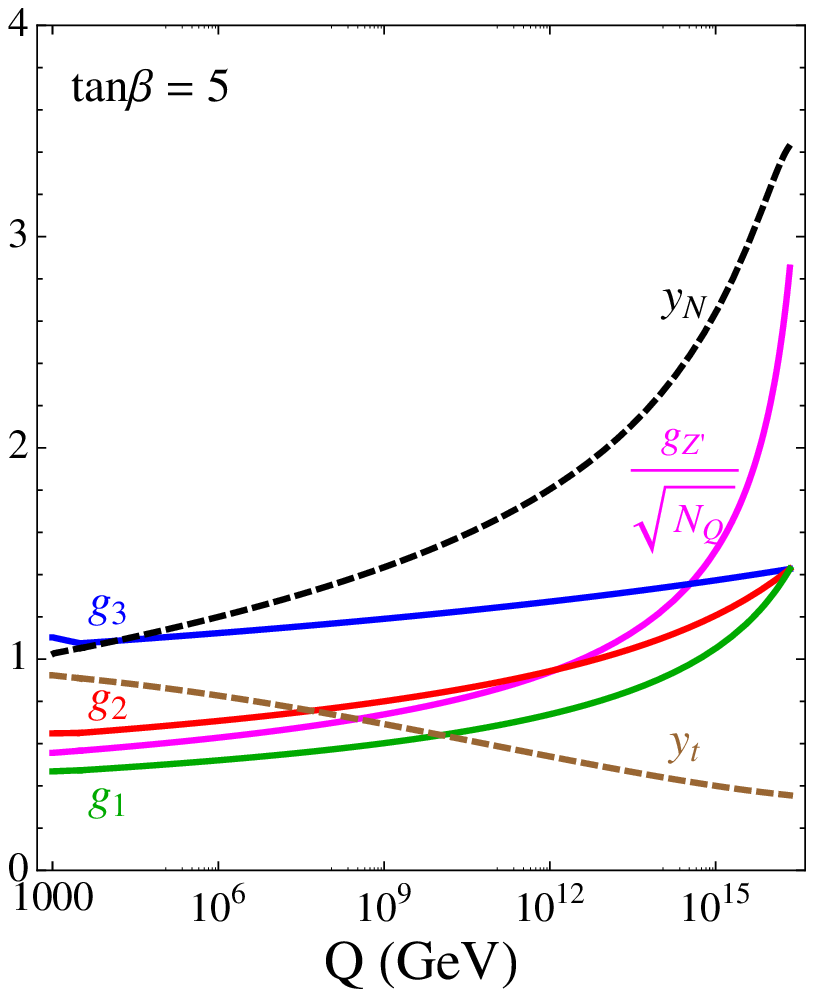}}
\hspace{0.2cm}
\subfigure[] 
{\includegraphics[width=0.51\linewidth]{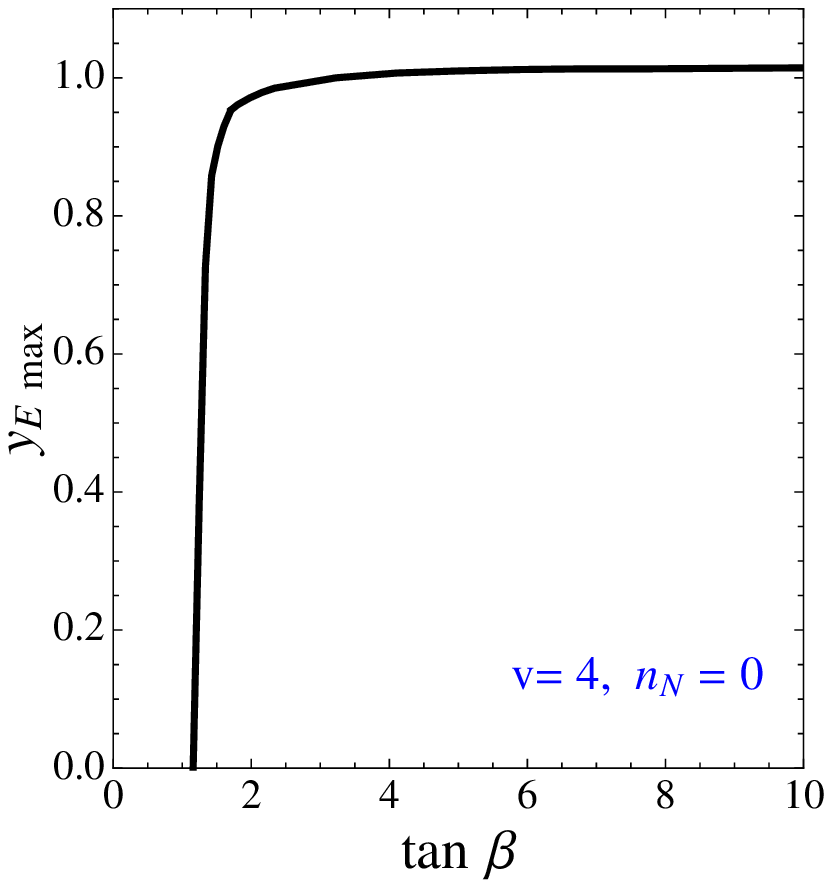}}
\end{center}
\caption{(a) RG runnings of the various couplings 
for the U(1)$_{Z'}$ extension (Case II).
The MSSM gauge couplings $\{g_3,g_2,g_1\}$ are unified, 
while $g_{Z'}/\sqrt{N_Q}$ is taken to be twice than the other gauge couplings at the GUT scale ($=2\times 10^{16}~{\rm GeV}$). 
(b) $y_{E \rm max}$ vs. tan$\beta$ for the U(1)$_{Z'}$ extension (Case II). Due to the U(1)$_{Z'}$ gauge interaction, the upper bound on $y_E$ at the EW scale, $y_{E \rm max}$ is lifted from $0.7$ to $1.0$ at ${\rm tan}\beta=2$.  
}
\end{figure}

In Case II, the LP constraint on $|y_E|$ is relaxed up to $1.02$ for $v=4$, $n_N=0$, and ${\rm tan}\beta\gtrsim 2$. 
It reaches the perturbativity bound ($\approx 3.4$) at the GUT scale. 
For $|y_E|\gtrsim 1$, thus, the model becomes strongly-coupled at the GUT scale. See Figure 2. 
Here the masses of $\{Q,Q^c;U,U^c;D,D^c\}$ are set to be 3 TeV. 
Hence, \eq{requirement} is fulfilled for large ${\rm tan}\beta$ ($\gtrsim 10$).
In this case, $h^0\rightarrow \gamma\gamma$ can be  enhanced more naturally than the case without the gauged U(1)$_{Z'}$. 
Since the U(1)$_{Z'}$ normalization is not determined, 
we cannot discuss the gauge coupling unification including $g_{Z'}$. In Figure 2, $g_{Z'}/\sqrt{N_Q}$
is set to be twice than the other gauge couplings at the GUT scale. 
We naively suppose that a proper (higher dimensional) UV theory at the GUT scale determines the normalization such that all the gauge couplings are unified. 
Unlike Case I, the maximal value of $|y_E|^4$ at low energy is just around unity in Case II. Hence, the oblique parameters can safely reside in the 1$\sigma$ band for $|y_E|\sim 1$, $\mu_L\gtrsim 210$ GeV, and ${\rm tan\beta}\gtrsim 10$.

\section{Conclusion}
\label{sec:conclusion}

From the effective potential approach, we have seen 
that the vector-like leptons $\{L,L^c;N,N^c\}$ can efficiently enhance the radiative correction to the Higgs mass,  
if their relevant Yukawa coupling to the MSSM Higgs, $y_N$ is of order unity. 
Without assuming the large mixing of the stops, thus, 
the 125 GeV Higgs mass can be explained 
with $\widetilde{m}_t^2\sim (500~{\rm GeV})^2$. 
Since the mass bounds on extra leptons are not so stringent yet compared to those of colored particles, 
our assumption of their relative small masses can make it possible to avoid the fine-tuning in the Higgs sector. 

In order to maintain the perturbativity of $y_N$ also  at higher energy scales, 
we introduced a new gauge symmetry,
under which only the extra vector-like leptons $\{L,L^c;N,N^c\}$ are charged but all the MSSM superfields including the Higgs remain neutral. 
As a simple example, we proposed an SU(2)$_{Z'}$ gauge extension, 
in which all the gauge couplings including that of SU(2)$_{Z'}$ can be unified at the GUT scale. 
In this case, the maximal value of $y_N$ allowed at low energy is lifted up to 1.78 (2.0) for ${\rm tan}\beta\gtrsim 3$ depending on the matter contents,  
which is enough to explain the 125 GeV Higgs mass. 

If the charged leptons $\{E,E^c\}$ are introduced instead of $\{N,N^c\}$, 
one can apply the same idea for enhancing di-photon decay rate of the Higgs. 
For the perturbativity and the unification of the MSSM gauge couplings, however, only a U(1)$_{Z'}$ gauge extension is possible in this case. 
By the U(1)$_{Z'}$ gauge interaction, the LP constraint on $y_E$ is relaxed up to $y_E\lesssim 1.02$ 
for ${\rm tan}\beta\gtrsim 2$ at the EW scale. 
In this case a relatively larger ${\rm tan}\beta$ ($\gtrsim 10$) is preferred for explaining 125 GeV Higgs mass. 

\acknowledgments
 
This research is supported by Basic Science Research Program through the 
National Research Foundation of Korea (NRF) funded by the Ministry of Education, Grant No. 2013R1A1A2006904 (B.K.) and 2011-0011083 (C.S.S.), and also in part by Korea Institute for Advanced Study (KIAS) grant funded by the Korea government 
(B.K.).
C.S.S acknowledges the Max Planck Society (MPG), the Korea Ministry of Education, Science and Technology (MEST), Gyeongsangbuk-Do and Pohang City for the support of the Independent Junior Research Group at the Asia Pacific Center for Theoretical Physics (APCTP).


\end{document}